\documentclass[9pt,twocolumn,twoside]{osajnl}
\usepackage{algorithm}
\usepackage{algpseudocode}
\usepackage{physics}

\journal{jocn}

\usepackage[justification=justified,singlelinecheck=false]{caption}

\usepackage{xcolor}
\usepackage[normalem]{ulem}

\definecolor{addedblue}{RGB}{0,80,180}
\definecolor{deletedred}{RGB}{180,0,0}
\renewcommand{\thesubsection}{\thesection.\Alph{subsection}}

\setboolean{shortarticle}{false}

\title{A Framework for Quantum Data Center Emulation Using Digital Quantum Computers}

\author[1]{Seyed Navid Elyasi}
\author[1]{Paolo Monti}
\author[2]{Jun Li}
\author[1, *]{Rui Lin}

\affil[1]{Department of Electrical Engineering, Chalmers University of Technology, Gothenburg, Sweden}
\affil[2]{School of Electrical and Information Engineering, Soochow University, China}
\affil[*]{elyasi@chalmers.se}

\begin{abstract}
As quantum computers scale, single-chip architectures face inherent limitations in qubit count. This drives the need for modular quantum computing and Quantum Data Centers (QDCs), where multiple quantum processor units (QPUs) are interconnected to enable the distributed execution of a quantum algorithm. However, evaluating distributed quantum computing (DQC) architectures is challenging.  Classical simulation is limited by the exponential growth of the state vector, limiting its ability to model large systems and realistically capture hardware noise and timing. Meanwhile, implementing QDC introduces interconnect noise challenges such as transduction inefficiency and optical fiber loss. In this work, we introduce a hardware-based emulation framework by partitioning a single quantum processor's qubit coupling map into multiple logical QPUs. We show how noise arising from transduction and optical fiber can be modeled by adding an ancilla qubit representing the environment,  based on quantum collisional dynamics. This model is then translated into a gate-based circuit, in which the couplings between each portion act as controllable noisy quantum communication channels. We demonstrate the framework on IBM quantum hardware by executing remote gates under controllable communication noise. To highlight the flexibility of the platform, we further replicate the implementation results of distributed Grover's search algorithm on an ion-trap system. Finally, we test a larger circuit, i.e., a five-qubit Quantum Fourier Transform (QFT), achieving reasonable fidelity across logical QPUs. Overall, the framework provides a scalable hardware-level emulation platform that captures noise sources through physical qubits, and is compatible with any platform supporting the Qiskit SDK.
\end{abstract}

\setboolean{displaycopyright}{false}

\begin{document}

\maketitle

% ------------------------------------------------------------
%  Section 1 – Introduction
% ------------------------------------------------------------
\section{Introduction}
\label{sec:Introduction}

\begin{figure}[t!]
\centering
\includegraphics[width=\columnwidth]{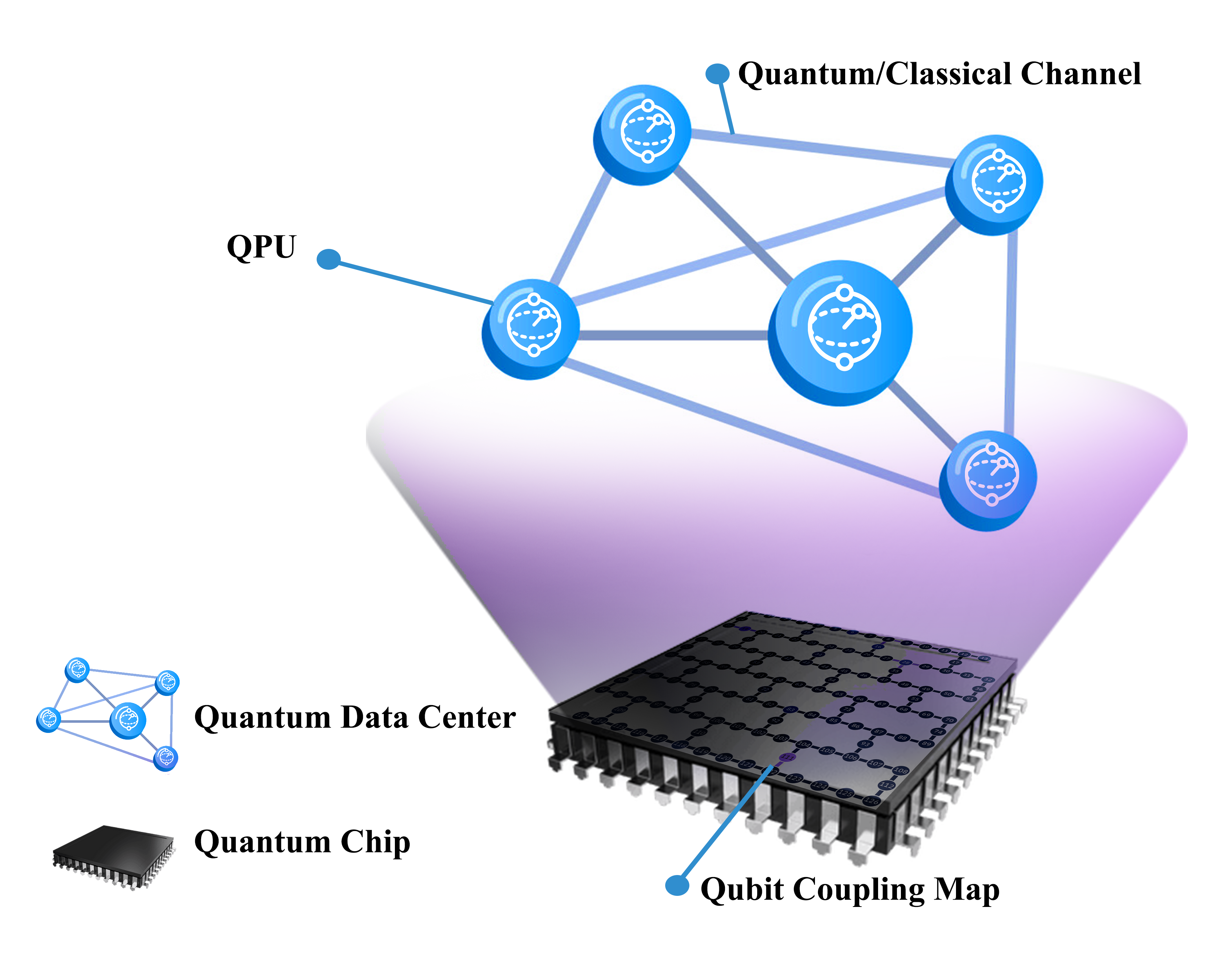}
\caption{A visual illustration of the core concept: a single quantum chip emulates a network of interconnected QPUs. The chip's coupling map is partitioned, with each partition considered as a logical QPU, and the existing couplings between groups interpreted as virtual interconnects between them.}
\label{f:Demo}
\end{figure}

Current quantum computing platforms remain limited in the number of qubits that can be reliably integrated on a single processor~\cite{chae2024quantum_qubit_review, campbell2024quantum, shapourian2025quantum, sebastiano2019scalable}. Such limitations have led to the concept of Quantum Data Centers (QDCs), in which multiple quantum processor units (QPUs) are interconnected to function as a single distributed system~\cite{liu2024quantum, shapourian2025quantum, cacciapuoti2025datacenters}. QDCs are being explored for a range of potential applications, including quantum sensing, magic state distillation, and distributed quantum computing (DQC)~\cite{liu2024quantum}. Among these, DQC has gained particular attention, both as a scalable solution to the qubit integration challenge and as a strategic priority in the roadmaps of major players in the field~\cite{campbell2024quantum, divincenzo2025thirty}.

\begin{figure*}[t!]
\centering
\includegraphics[width=1\textwidth]{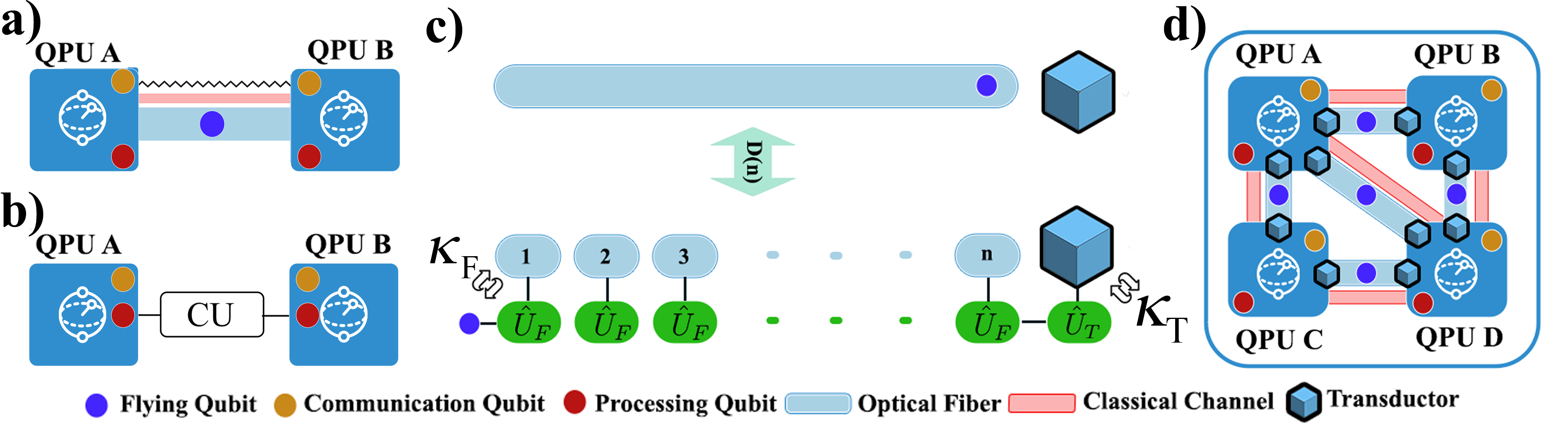}
\captionsetup{width=1\textwidth}
\caption{(a) Interconnected QPUs within a QDC. (b) Non-local operations between interconnected QPUs, where CU denotes a controlled-unitary gate. (c) Representation of the noise model, in which a flying qubit interacts with its surrounding environment, including the optical fiber and transducer, resulting in continuous information leakage. Using a collision model (CM), the optical fiber and transducer environments are discretized into independent segments that sequentially interact with the system via unitary operators $\hat{U}_F$ (fiber) and $\hat{U}_T$ (transducer), with coupling strengths $\kappa_F$ and $\kappa_T$, respectively. (d) Example QDC in which four QPUs (A--D) are interconnected in a mesh topology via optical
fibers.}
\label{f:DQCarch}
\end{figure*}

DQC relies on both quantum and classical communication channels to coordinate the execution of algorithms across multiple QPUs~\cite{hwang2024machine, cacciapuoti2025datacenters}. Each QPU is responsible for a portion of the computing task, and remote gates (RGs), such as CNOT gates or more general controlled-unitary (CU) gates, are used to perform logical operations between QPUs~\cite{caleffi2024distributed, peckham2024async, gottesman1999teleportation}. Within each QPU, qubits are categorized by their distinct roles (see Fig.~\ref{f:DQCarch}): processing qubits handle the local part of the quantum algorithm, communication qubits act as the interface for establishing entanglement between QPUs, and flying qubits (FQs) serve as quantum information carriers, typically implemented via photons, to physically transmit quantum information across the network. This modular structure enables DQC to support scalable computation but also introduces challenges~\cite{ferrari2023modular, vanMeter2022distributed}.

Various experimental efforts have been undertaken to realize QDCs across different hardware platforms~\cite{ritter2012elementary, hofmann2012heralded,  krutyanskiy2023entanglement,zhou2026photonic}. For example, ion traps (with a 2 m distance from each other) have been connected via SMF-type fibers with a 422 nm wavelength~\cite{main2025distributed}, superconducting qubits have been linked using 6 cm NbTi cables~\cite{almanakly2025chiral}, and photonic systems have employed 850 nm interconnects~\cite{aghaeerad2025scaling}. However, these implementations remain in their early stages, typically involving only a few qubits and exhibiting low interconnection efficiency~\cite{barral2025review}. Consequently, practical and accessible testbeds for investigating QDC architectures remain limited.

Theoretical research efforts in DQC have been focused on addressing major challenges such as interconnect noise, RG fidelity, and system scalability~\cite{liu2025hardware, campbell2024quantum, elyasi2025toward}. However, performing a high-fidelity CU gate between distant QPUs remains technically challenging due to the difficulty of maintaining entanglement over noisy links~\cite{barral2025review, shapourian2024_scalable_entanglement}. Several simulation frameworks have been developed to study DQC and quantum networking systems. For example, tools such as NetSquid~\cite{coopmans2021netsquid} and SeQUeNCe~\cite{wu2021sequence} provide powerful classical simulation tools for modeling quantum networks, communication protocols, and entanglement distribution.

However, these frameworks rely on classical simulation and abstract noise models, which cannot fully capture hardware-specific effects such as device calibration, native gate errors, connectivity constraints, and execution timing on real quantum processors. Consequently, there remains a lack of practical experimental platforms for studying DQC architectures under realistic hardware conditions.

Quantum processors can be used to emulate quantum systems under realistic hardware conditions~\cite{cao2023abinitio, Karimi2024, grimsley2019adaptive, fauseweh2024quantum, elyasi2025three}. Unlike in classical simulation, quantum processors operate with physical qubits and therefore naturally incorporate device-level noise and operational constraints~\cite{ibm2023quantum}. Moreover, classical state vector simulations scale exponentially with system size and cannot efficiently capture hardware-level timing effects and correlated noise processes at scale~\cite{feynman1982simulating}. Therefore, quantum computers can also be leveraged to emulate the behavior of quantum networks, specifically QDCs~\cite{riera-sabat2025quantum}.

In this paper, we present a hardware-based emulation framework for studying DQC architectures using a single quantum processor. As shown in Fig.~\ref{f:Demo}, our approach partitions the coupling map of a QPU into multiple virtual QPUs, each representing a logically distinct node, and emulates communication channels between them to model noisy inter-QPU interactions. To represent communication-induced impairments, we incorporate a quantum collision model (CM) in which ancillary qubits interact sequentially with system qubits and are periodically reset to emulate environmental noise effects on the communication channel.

We implement the proposed framework on IBM quantum hardware and emulate RG operations under controllable communication noise. These emulations demonstrate the impact of communication-induced noise on interconnected QPUs. Using the emulated RGs, we further implement distributed versions of Grover's search algorithm~\cite{grover1996fast}, whose behavior is compared with previously reported experimental results with distributed ion-trap systems~\cite{main2025distributed}. We then extend this to a more complex scenario by using the emulator to run a five-qubit Quantum Fourier Transform (QFT) and apply tomography, thereby illustrating the practical capability of our framework for complex DQC scenarios.

The remainder of this paper is organized as follows: 
Section~\ref{sec:SystemModel} presents the system model and 
theoretical foundations. Specifically, Section~\ref{sec:II-A} 
introduces the concept of RGs, while Section~\ref{sec:II-B} 
details the theoretical foundations of the CM and its 
integration into our framework. Section~\ref{sec:II-C} 
then describes the proposed interconnected system and the 
algorithms used to implement the emulation approach. 
Section~\ref{sec:Results} presents the emulation results, 
beginning with the experimental setup in 
Section~\ref{sec:setup}, followed by the execution of 
various RG protocols in Section~\ref{sec:RemoteGatesExecution}. 
The analysis is then extended to more complex scenarios, 
demonstrating the emulation of distributed Grover's search algorithm
and the QFT in Section~\ref{sec:Grover} 
and Section~\ref{sec:QFT}, respectively. Finally, 
Section~\ref{sec:Conclusion} concludes the paper.

% ------------------------------------------------------------
%  Section 2 – System and Model
% ------------------------------------------------------------
\section{System and Model}
\label{sec:SystemModel}

% ── 2.1 Remote Gate ──────────────────────────────────────────
\subsection{Remote Gate (RG)}
\label{sec:II-A}

RGs are essential for executing quantum algorithms distributively that rely on strong correlations between processing qubits, such as the QFT and Grover's search algorithm~\cite{grover1996fast, shor1994algorithms}. A basic example is shown in Fig.~\ref{f:DQCarch}(b), where QPU A's processing qubit (red circle) acts as the control and QPU B's processing qubit serves as the target, with a CU operation applied between them. Both qubits are directly involved in the logic of the distributed algorithm.

Unlike local gates, which operate within a single QPU, RGs require both quantum entanglement, as a quantum channel, and a classical communication channel for feed-forward operations that coordinate the outcomes of mid-circuit measurements. These channels are illustrated as blue and red rectangles between QPUs in Fig.~\ref{f:DQCarch}, respectively. Several protocols have been proposed to implement such gates, including cat-state communication-based (Cat-Comm), single teleportation (TP1), double teleportation (TP2), and TP-Safe~\cite{campbell2024quantum}.

In the Cat-Comm protocol, a remote CNOT gate is applied between the control qubit $q_{1}^{A}$ on QPU A and the target qubit $q_{4}^{B}$ on QPU B, as illustrated in Fig.~\ref{f:SimulateGates}(a). The protocol begins with a Cat-entangler operation, which uses a Bell pair shared between the communication qubits of the two QPUs ($q_{3}^{A}$ in QPU A and $q_{1}^{B}$ in QPU B) to temporarily share the control qubit's state with the communication qubit on QPU B. Once this distributed control state has been established, a local CNOT gate is applied between the communication qubit $q_{1}^{B}$ and the target qubit $q_{4}^{B}$. Finally, a Cat-disentangler operation removes the temporary distributed correlation and releases the communication qubits for future RG operations. Consequently, Cat-Comm has three important characteristics: it requires only a single Bell pair, the communication qubits are released after the operation, and the state of the original control qubit $q_{1}^{A}$ remains unchanged throughout the protocol.

In contrast, the TP1 protocol, shown in Fig.~\ref{f:SimulateGates}(b), realizes the remote CNOT by teleporting the state of the control qubit $q_{1}^{A}$ to the communication qubit $q_{1}^{B}$ on QPU B using a Bell pair shared between the communication qubits $q_{3}^{A}$ and $q_{1}^{B}$. Unlike Cat-Comm, the control qubit's state is not shared but teleported to the destination QPU. After teleportation, the communication qubit on QPU B acts as the control for a local CNOT gate applied to the target qubit $q_{4}^{B}$. As a result, TP1 has two main limitations: the destination communication qubit remains occupied after the operation and the logical state is displaced from its original location, meaning that it must be teleported back if subsequent computation requires the original qubit placement.

To address these limitations, TP2 and TP-Safe were proposed (Fig.~2 of reference.~\cite{campbell2024quantum}). All three teleportation-based protocols (TP1, TP2, and TP-Safe) share the same initial teleportation procedure. However, TP2 performs an additional teleportation after the RG has been executed, returning the logical state to QPU A and releasing the communication qubit on QPU B. TP-Safe extends this approach further by restoring the logical state to its original processing-qubit location, typically through a SWAP operation, thereby preserving both the logical qubit placement and communication-qubit availability. These improvements come at the cost of additional resource consumption, since the return teleportation requires an extra Bell pair, additional measurements, classical communication, and correction operations, all of which increase circuit depth and the overall error budget.

The quality of the entangled states shared between QPUs is a critical performance factor for RGs. Ideally, such gates rely on high-fidelity Bell pairs (e.g., $\ket{\Phi^+}$) established between communication qubits, which are depicted as orange circles in Fig.~\ref{f:DQCarch}(a), with entanglement illustrated as zigzag lines. However, in practice, these states are often degraded by noise due to communication and hardware inefficiencies such as QPU coherence time, fiber loss, transducer conversion rate, and overall noise. Recent studies have modeled these effects using both stochastic simulations and circuit-level approaches to understand fidelity degradation and to guide the design of more robust RG implementations across distributed QPUs~\cite{ferrari2023modular, campbell2024quantum}.

 In this work, Cat-Comm is used as the primary RG protocol because it requires only a single shared Bell pair and avoids the state-restoration procedures required by TP1, TP2, and TP-Safe. Consequently, it incurs lower communication, entanglement-distribution, and local-gate overhead, making it an attractive choice for near-term DQC systems. TP1 is additionally included because it captures the core teleportation mechanism common to TP2 and TP-Safe, while demonstrating the ability of the proposed framework to support multiple RG protocols beyond Cat-Comm.

% ── 2.2 Collisional Model ────────────────────────────────────
\subsection{Collisional Model (CM)}
\label{sec:II-B}

To emulate communication noise in interconnected QPUs within a QDC, we adopt a quantum collision model (CM) framework~\cite{Ciccarello2022, elyasi2024experimental}. CMs are widely used to describe open quantum system dynamics through repeated interactions between a system and a sequence of environmental ancillas~\cite{ziman2004collision, cattaneo2022_brief_journey_collision_models}.

This approach allows us to capture the degradation of shared entanglement and the effective loss of connectivity between remote QPUs. The procedure used to incorporate these noise processes within the emulation framework is summarized in Algorithm~\ref{alg:qdc_framework}. In the following, we describe the main components of the model and their implementation, following the structure and notation defined in the algorithm.

As depicted in Fig.~\ref{f:DQCarch}(c), a FQ mediates entanglement between communication qubits while interacting with its surrounding environment, such as optical fibers and transducer interfaces. These interactions are inherently non-unitary and irreversible, as information leaks from the FQ into the environment at rates proportional to the coupling strengths $\kappa_F$ (fiber) and $\kappa_T$ (transducer).

Under the Born approximation~\cite{breuer2002theory}, which assumes that the FQ and the environment are initially uncorrelated, the joint system–environment state evolves unitarily according to $\rho_{FQ,E}(t) = U(t)\,\rho_{FQ}(0)\otimes\rho_E(0)\,U^\dagger(t),$
where $U(t)=e^{-i H t}$ (with $\hbar=1$) is generated by the interaction Hamiltonian $H$. The reduced state of the FQ is obtained by taking the partial trace, $\rho_{FQ}(t)=\operatorname{Tr}_E\!\left[\rho_{FQ,E}(t)\right].$ This procedure captures the effective loss of coherence experienced by the FQ as information leaks into the environment during the interaction.

In the CM description~\cite{ziman2004collision,ziman2010simple,Ciccarello2022, elyasi2024experimental}, the environment is represented as a sequence of independent ancilla systems that interact with the FQ in a series of discrete steps. Each interaction, or ``collision'', corresponds to a short coupling event between the FQ and a local environment representing a segment of the communication channel. Denoting the environment ancillas by $\{E_1,\dots,E_n\}$ and the corresponding unitaries by $\{U_1,\dots,U_n\}$, the global state after $n$ collisions is
\begin{equation}
\label{eq:dynamics}
\rho_n = U_n\cdots U_1 \bigl(\rho_{FQ}\otimes\rho_E\bigr) U_1^\dagger\cdots U_n^\dagger,
\end{equation}
and the reduced FQ state becomes  $\rho^{FQ}_n = \operatorname{Tr}_{E_1,\dots,E_n}\!\left[\rho_n\right].$ This formulation models the progressive degradation of the flying-qubit state as it interacts sequentially with different segments of the communication channel.

In our implementation, each collision is described by a unitary interaction between the FQ and an environment ancilla. Each interaction $U_j = e^{-i H_j \Delta t}$ is generated by an amplitude-damping Hamiltonian
\begin{equation}
\label{Eq:Hamiltonian}
H_j = \kappa \bigl(\sigma_+^{FQ}\!\otimes\!\sigma_-^{E_j} + \sigma_-^{FQ}\!\otimes\!\sigma_+^{E_j}\bigr),
\end{equation}
where $\kappa$ denotes a tunable coupling constant (potentially distinct for fiber and transducer components), and $\sigma_\pm$ are the raising and lowering operators. This interaction corresponds to an excitation-exchange process between the FQ and the environment ancilla and therefore models energy relaxation and information leakage into the surrounding environment.

In the Markovian limit, the reduced dynamics in the interaction picture (with respect to the free environmental Hamiltonian $H_j$) satisfy the master equation
$ \frac{d\rho_S}{dt} = \kappa\,\mathcal{D}[\sigma_-]\rho_S,$ with Lindblad dissipator defined as $\mathcal{D}[O]\rho = O\rho O^\dagger - \tfrac{1}{2}\!\left(O^\dagger O\rho + \rho O^\dagger O\right)$.

Establishing high-fidelity entanglement between communication qubits of distinct QPUs is a central requirement. Ideally, as shown in Fig.~\ref{f:DQCarch}(a), the communication qubits $q_{\mathrm{comm}}^{A}$ and $q_{\mathrm{comm}}^{B}$ share the Bell state
$(\ket{\Phi^+}). $ In practice, however, channel imperfections and environmental coupling reduce coherence, yielding a mixed entangled state instead of the ideal Bell pair.

A practical advantage of the CM framework is that its dynamics are constructed from elementary unitary interactions, which can be mapped directly onto quantum circuits. This property enables experimental emulation of communication noise without requiring pulse-level control or additional hardware-specific approximations.

In this work, we apply the CM formalism to emulate communication noise in a network of superconducting QPUs. Although the framework is general and can be applied to different hardware platforms, our experimental implementation is demonstrated using IBM quantum processors, as described in the following section.

% ── 2.3 Interconnected System and Algorithms ─────────────────
\subsection{Interconnected System and Algorithms}
\label{sec:II-C}

\begin{figure}[t!]
\centering
\includegraphics[width=\columnwidth]{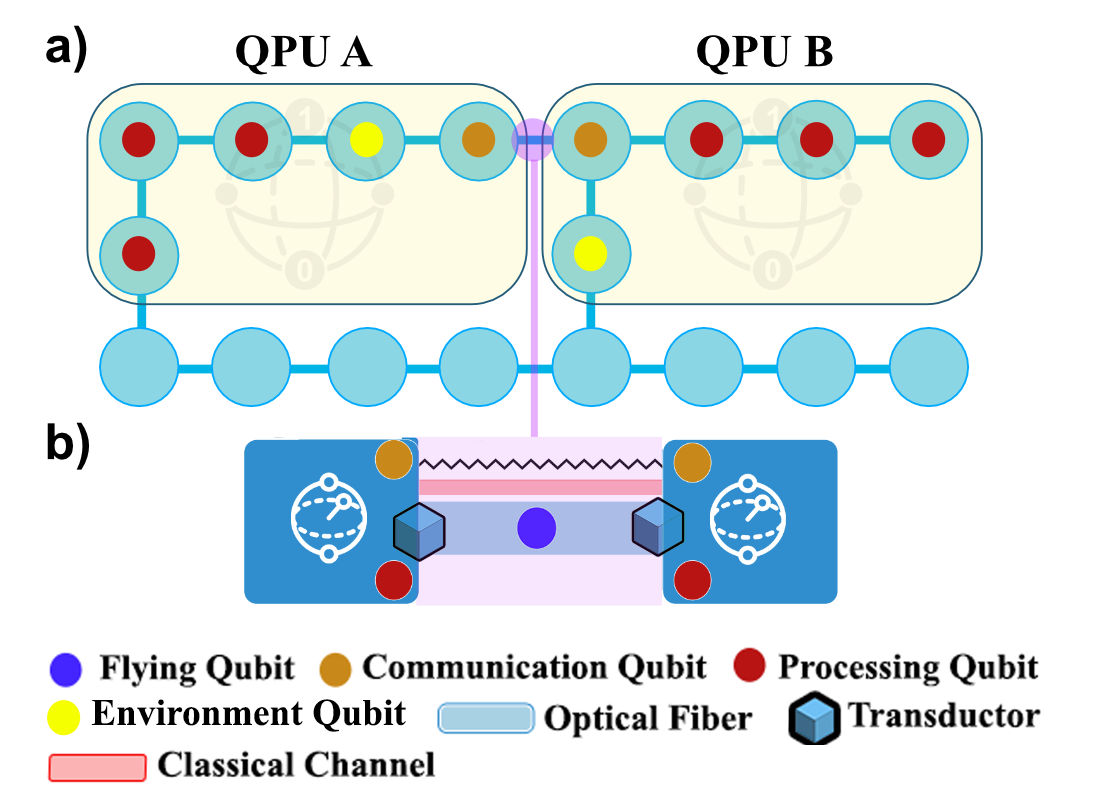}
\caption{(a) Illustrates the partitioning of a section of the qubit coupling map of a superconducting quantum computer into two distinct QPUs. (b) Highlights that the coupling between nearest-neighbor qubits, designated as communication qubits belonging to different QPUs, can be utilized as a quantum channel between the two processors.}
\label{f:MappingCircuit}
\end{figure}

Our objective is to emulate DQC within the context of a QDC. Although this framework can be applied to various scenarios, we consider an example in which multiple superconducting QPUs are interconnected in a mesh topology using optical fibers and transducers, as shown in Fig.~\ref{f:DQCarch}(d). Each QPU consists of: {\it (i)} flying qubits (blue circles), which physically carry quantum information between QPUs, {\it (ii)} communication qubits (orange circles), stationary qubits responsible for establishing remote entanglement, {\it (iii)} processing qubits (red circles), which execute the quantum logic operations, and {\it (iv)} transducers (blue cubes), which are coupled to communication qubits and convert microwave-frequency signals in the GHz range (typically 4--8\,GHz for superconducting qubits) to optical-frequency photons in the hundreds-of-THz range (typically $\sim$193\,THz for C-band telecom wavelengths) for inter-QPU transmission.

To emulate the QDC on the qubit coupling map of a superconducting quantum computer, the first step, as also indicated in Algorithm~\ref{alg:qdc_framework}, is to represent the device connectivity as a coupling graph $G=(V,E)$. In this graph, $V$ denotes the set of vertices corresponding to physical qubits, while $E$ represents the available couplings between them. As an example, shown in Fig.~\ref{f:MappingCircuit}(a), a portion of the coupling map of a superconducting quantum processor can be partitioned into two logical units, denoted as QPU A and QPU B. Each partition may contain processing qubits, communication qubits, and environment qubits depending on their role in the emulation framework. In our implementation, the ancilla (yellow circles) is only used during the collision-model noise step. It is therefore assigned to an available physical qubit close to the corresponding communication qubit, preferably with direct hardware connectivity as shown in Fig.~\ref{f:MappingCircuit}(a). Since this ancilla is reset and reused after each collision and is not involved in the remaining remote-gate protocol, one such neighboring qubit per QPU is sufficient. This requirement is not restrictive and becomes easier to satisfy on processors with denser connectivity, such as IBM’s Nighthawk architecture.
\begin{algorithm}[t!]
\caption{DQC Emulation on a Single QPU}
\label{alg:qdc_framework}
\begin{algorithmic}[1]
\Require $\mathcal{B}$ with coupling graph $G=(V,E)$, task $T$, distributed circuit $\mathcal{A}$, number of partitions $N$, collision parameters $(\kappa_T,\kappa_F,n)$, remote protocol $\mathcal{P}$
\Ensure $p_{\mathrm{succ}}$, $p(x)$, or $F$

\State Find $\{V_i\}_{i=1}^N$ such that
\Statex \hspace{\algorithmicindent} $V=\bigsqcup_{i=1}^N V_i,\quad V_i=Q_i^{\mathrm{proc}}\sqcup Q_i^{\mathrm{comm}}$
\Statex \hspace{\algorithmicindent} $(u,v)\in E,\ u\in V_i,\ v\in V_j,\ i\neq j \Longrightarrow u\in Q_i^{\mathrm{comm}},\ v\in Q_j^{\mathrm{comm}}$
\Statex \hspace{\algorithmicindent} $u\in Q_i^{\mathrm{comm}},\ v\in Q_j^{\mathrm{comm}},\ i\neq j \Longrightarrow \mathrm{dist}_G(u,v)\in\{1,\infty\}$

\State Map $\mathcal{A}\mapsto \{\mathcal{A}_i\}_{i=1}^N$

\ForAll{$(i,j)\in \mathcal{R}(\mathcal{A},T)$}
    \State $\rho \gets \ket{\Phi^+}\!\bra{\Phi^+}$, where $\ket{\Phi^+}=(\ket{00}+\ket{11})/\sqrt{2}$
    \State $\rho \gets \mathrm{Tr}_{\mathrm{env}}\!\left[U_T(\kappa_T)\bigl(\rho\otimes\ket{0}\!\bra{0}\bigr)U_T^\dagger(\kappa_T)\right]$
    \For{$k=1$ to $n$}
        \State $\rho \gets \mathrm{Tr}_{\mathrm{env}}\!\left[U_F(\kappa_F)\bigl(\rho\otimes\ket{0}\!\bra{0}\bigr)U_F^\dagger(\kappa_F)\right]$
    \EndFor
    \State $\rho_{ij}^{\mathrm{comm}} \gets \rho$
    \State $G_{ij} \gets \mathcal{P}(\rho_{ij}^{\mathrm{comm}})$
\EndFor

\If{$T=\textsc{RemoteGate}$}
    \State measure $p_{\mathrm{succ}}=\Pr(\mathrm{success})$
\ElsIf{$T=\textsc{QFT}$}
    \State execute $\mathcal{A}$ and obtain $\rho_{\mathrm{out}}$
    \State compute $F(\rho_{\mathrm{ideal}},\rho_{\mathrm{out}})=\left(\mathrm{Tr}\sqrt{\sqrt{\rho_{\mathrm{ideal}}}\rho_{\mathrm{out}}\sqrt{\rho_{\mathrm{ideal}}}}\right)^2$
\ElsIf{$T=\textsc{Grover's Search Algorithm}$}
    \State execute $\mathcal{A}$ and measure $p(x)=\Pr(x)$
\Else
    \State execute $\mathcal{A}$ and evaluate the task-specific metric
\EndIf
\end{algorithmic}
\end{algorithm}

An important requirement during this partitioning is the connectivity between different partitions. In particular, the connection between QPUs should not involve multi-hop paths through intermediate qubits, as illustrated in Fig.~\ref{f:MappingCircuit}(a). Instead, the coupling between the boundary qubits of two partitions directly represents the quantum channel between the corresponding QPUs, as pointed out in Fig.~\ref{f:MappingCircuit}(b). Formally, let the coupling graph be $G=(V,E)$ and let the partitions be $\{V_i\}_{i=1}^{N}$ such that
$$
V=\bigcup_{i=1}^{N} V_i , \qquad V_i \cap V_j = \varnothing \quad \forall i\neq j .
$$
Each partition is further divided into processing and communication qubits,
$$
V_i = Q_i^{\mathrm{proc}} \cup Q_i^{\mathrm{comm}}, \qquad
Q_i^{\mathrm{proc}} \cap Q_i^{\mathrm{comm}} = \varnothing .
$$
The boundary constraint requires that any inter-partition edge connects only communication qubits,
$$
(u,v)\in E,\; u\in V_i,\; v\in V_j,\; i\neq j
\;\Rightarrow\;
u\in Q_i^{\mathrm{comm}},\;
v\in Q_j^{\mathrm{comm}} .
$$
Moreover, to avoid multi-hop connections between QPUs, the distance between communication qubits belonging to different partitions must satisfy
$$
u\in Q_i^{\mathrm{comm}},\; v\in Q_j^{\mathrm{comm}},\; i\neq j
\;\Rightarrow\;
\mathrm{dist}_G(u,v)\in\{1,\infty\}.
$$
This condition ensures that any valid inter-QPU connection corresponds to a direct edge in the coupling graph. In other words, the boundary communication qubits act as the endpoints of the quantum channel between QPUs. If additional intermediate qubits were present between the partitions, they could instead be interpreted as quantum memories or elements of a quantum repeater, which would introduce additional resources and complexity into the model.

After mapping the QDC architecture onto the coupling graph, the next step is to model the relevant noise sources. As described in Algorithm~\ref{alg:qdc_framework}, we adopt the CM formalism, where noise is represented as sequential interactions between system qubits and environmental degrees of freedom. This setup is illustrated in Fig.~\ref{f:DQCarch}(c). In our model, FQs act as mediators that distribute entanglement between communication qubits belonging to neighboring QPUs. The quality of the generated Bell state $\ket{\Phi^+}$ therefore depends on how reliably the FQs are generated, transmitted, and converted between the microwave-frequency (GHz) and optical-frequency (hundreds of THz) domains. 

\begin{figure*}[!t]
\centering
\includegraphics[width=\textwidth]{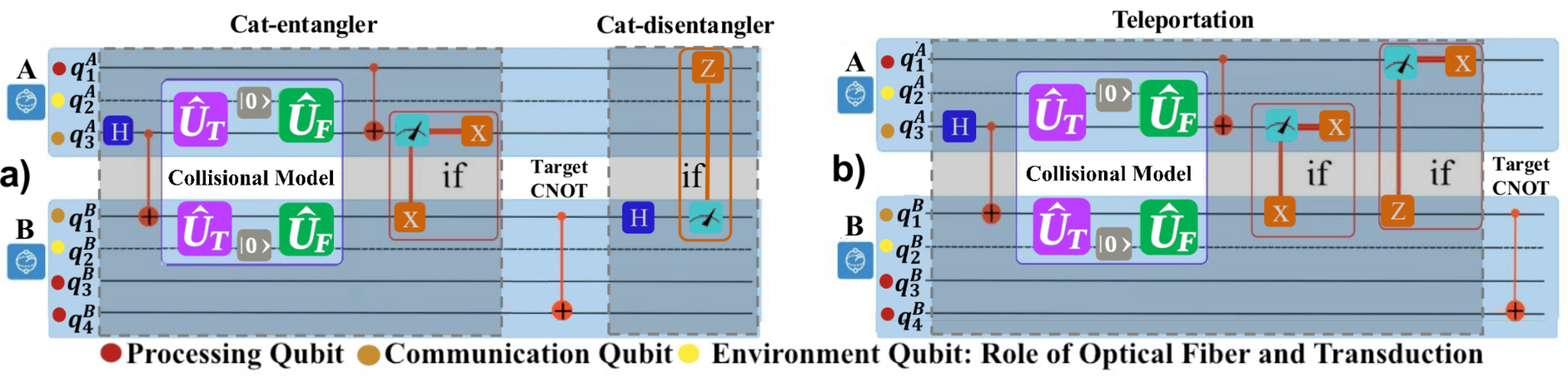}
\caption{Implementation of a remote CNOT gate using controllable noisy entanglement using the proposed collisional model: (a) Noisy cat-state communication (Cat-Comm)-based remote CNOT gate; (b) Noisy teleportation-based remote CNOT gate. The environment qubit, represented by the dashed line in the circuit, is not originally included in either of the illustrated protocols as shown in reference.~\cite{campbell2024quantum}. The grey boxes containing $\ket{0}$ represent reset gates, the purple gates denoted by $\hat{U}_{T}$ represent transduction operations, and the green gates denoted by $\hat{U}_{F}$ represent the optical-fiber-driven gates derived from the collisional model.}
\label{f:SimulateGates}
\end{figure*}

Since the RG protocols rely on the distribution of an entangled Bell pair between QPUs, the accumulated attenuation directly affects the quality of the shared entanglement. Starting from the ideal Bell state $\ket{\Phi^+}=(\ket{00}+\ket{11})/\sqrt{2}$, each collision induces an amplitude-damping interaction with coupling strength $\kappa$ as indicated in Eq.~\ref{Eq:Hamiltonian}. The excitation survival amplitude after one collision using Eq.~\ref{eq:dynamics} is $\cos(\kappa)$, giving a transmissivity of $\cos^2(\kappa)$. After $n$ independent collisions, the transmissivity becomes $\eta(n)=\cos^{2n}(\kappa)$. For the two-sided attenuation considered here, the Bell-pair fidelity with respect to the ideal Bell state is
\begin{equation}
\label{eq:fidelity}
F_{\mathrm{Bell}}(n)=\frac{1+\eta^2(n)}{2}.
\end{equation}
This expression directly relates the coupling strength and number of collision steps to the degradation of the distributed entanglement resource.

Quantum transduction remains one of the primary experimental challenges in DQC and is therefore modeled using a stronger coupling parameter than optical-fiber transmission. As a result,  in Eq.~\ref{Eq:Hamiltonian}, the $\kappa$ for the transduction and fiber coupling strengths could be denoted by $\kappa_{\mathrm{T}}$ and $\kappa_{\mathrm{F}}$, respectively, with $\kappa_{\mathrm{T}}>\kappa_{\mathrm{F}}$ to reflect the comparatively higher noise associated with current transduction technologies. Using $\kappa_{\mathrm{T}}=0.5$ gives $\eta=\cos^2(0.5)$ and therefore 
$F_{\mathrm{Bell}}\approx0.797$, which is in close agreement with the 
experimental value of $F=0.794^{+0.048}_{-0.071}$ 
reported by Meesala \textit{et al.}~\cite{meesala2024}, who demonstrated 
microwave--optical entanglement using a chip-scale piezo-optomechanical 
transducer, independently validating the physical basis of the chosen 
$\kappa_{\mathrm{T}}$ value. This value is used as an effective emulator setting to represent a noisy transduction stage while still allowing successful execution of the distributed protocols considered in this work.

In contrast, optical fibers exhibit relatively low loss over the short distances considered here; therefore, $\kappa_{\mathrm{F}}\ll1$, representing a weak system-environment interaction. Under this condition, $\cos(\kappa_{\mathrm{F}})\approx1-\kappa_{\mathrm{F}}^2/2$, which gives $\cos^2(\kappa_{\mathrm{F}})\approx1-\kappa_{\mathrm{F}}^2\approx e^{-\kappa_{\mathrm{F}}^2}$. Consequently, $\eta(n)=\cos^{2n}(\kappa_{\mathrm{F}})\approx e^{-n\kappa_{\mathrm{F}}^2}$, reproducing the expected exponential attenuation behavior of an optical communication channel. Using $D(n)=\gamma n/\alpha$ with $\gamma=\kappa_{\mathrm{F}}^2$ gives $\alpha D(n)=n\kappa_{\mathrm{F}}^2$, establishing the equivalence between the collision dynamics and physical fiber attenuation. 

For the fibers considered here, G-652-D, G-654-E, and G-655-D have attenuation coefficients $\alpha=0.0415$, $0.0392$, and $0.0507~\mathrm{km}^{-1}$, respectively. In this work, each collision step is used as a discrete emulator representation of a $D(1) = 10~m$ fiber segment in the emulation and gives $\alpha D(n)=n\kappa_{\mathrm{F}}^2$ to be $\alpha 0.01=\kappa_{\mathrm{F}}^2$ so that the $\kappa_F = \sqrt{0.01\alpha}$ can be computed and imported in the emulation. Therefore, increasing the number of collision steps corresponds to increasing the effective propagation distance, which in turn increases the accumulated channel attenuation.

With the system model established and the algorithms defined for both the emulation framework and the mapping function, the next step is to address the implementation of remote quantum gates within the simulation framework. In Algorithm~\ref{alg:qdc_framework}, these operations are represented as tasks $T$. Implementing such remote operations on real quantum hardware allows us to evaluate how the practical behavior of the system deviates from the ideal theoretical model. In particular, this step enables us to assess the impact of hardware noise, limited connectivity, and imperfect entanglement distribution on the performance of distributed quantum operations.

% ------------------------------------------------------------
%  Section 3 – Results and Discussion
% ------------------------------------------------------------
\section{Results and Discussion}
\label{sec:Results}

% ── 3.1 Experimental Setup ───────────────────────────────────
% NOTE: Added per Reviewer 2 reproducibility comment
\subsection{Experimental Setup}
\label{sec:setup}
All hardware experiments are conducted on IBM's \texttt{ibm-torino} backend, a 133-qubit Heron~R1 processor. Quantum circuits are implemented using the Qiskit SDK~\cite{javadi2024quantum} and transpiled with optimization level~3 using the native basis gate set $\{$CZ, I, RZ, SX, X$\}$. The CM time-evolution operators are generated using the QuTiP Python package~\cite{johansson2013qutip2}, while all DQC workflows, RG implementations, and noise-model emulations are executed within the proposed framework. Each experimental data point is obtained from 10{,}000 measurement shots. To ensure reproducibility, the GitHub repository~\cite{nelyasi_QDC} contains the source code, circuit implementations, transpilation settings, backend configuration information, hardware calibration data, job JSON files, selected qubits, and the corresponding device metrics, including $T_1$ and $T_2$ coherence times, two-qubit gate error rates, and readout error rates associated with the experiments reported in this work.

Classical feed-forward is required only for the correction stage of the Cat-Comm protocol and is not used within the CM emulation itself. Numerical reference results are obtained using Qiskit's AerSimulator, which is a CPU-based classical simulator, configured using the \texttt{from\_backend()} routine to construct a noise model from the calibration data of the target IBM Torino processor. This enables a direct comparison between hardware execution and calibration-informed simulation, allowing the impact of device-specific noise and runtime hardware variations on DQC performance to be evaluated.

% ── 3.2 Remote Gate Execution ────────────────────────────────
\subsection{RGs Execution}
\label{sec:RemoteGatesExecution}

Following the interconnected system introduced in Section~\ref{sec:II-C}, we demonstrate how to implement our proposed model in Section~\ref{sec:II-A}, enabling fine-tunable entanglement generation and, ultimately, the execution of RGs through a noisy, controllable quantum communication channel. While the focus here is on the CNOT gate, this approach can be generalized to any CU operation, as the CNOT is a special case where the target gate is the Pauli-X operator. This generality makes the method applicable to a wide range of quantum algorithms.

\begin{figure*}[t!]
\centering
\includegraphics[width=1\textwidth]{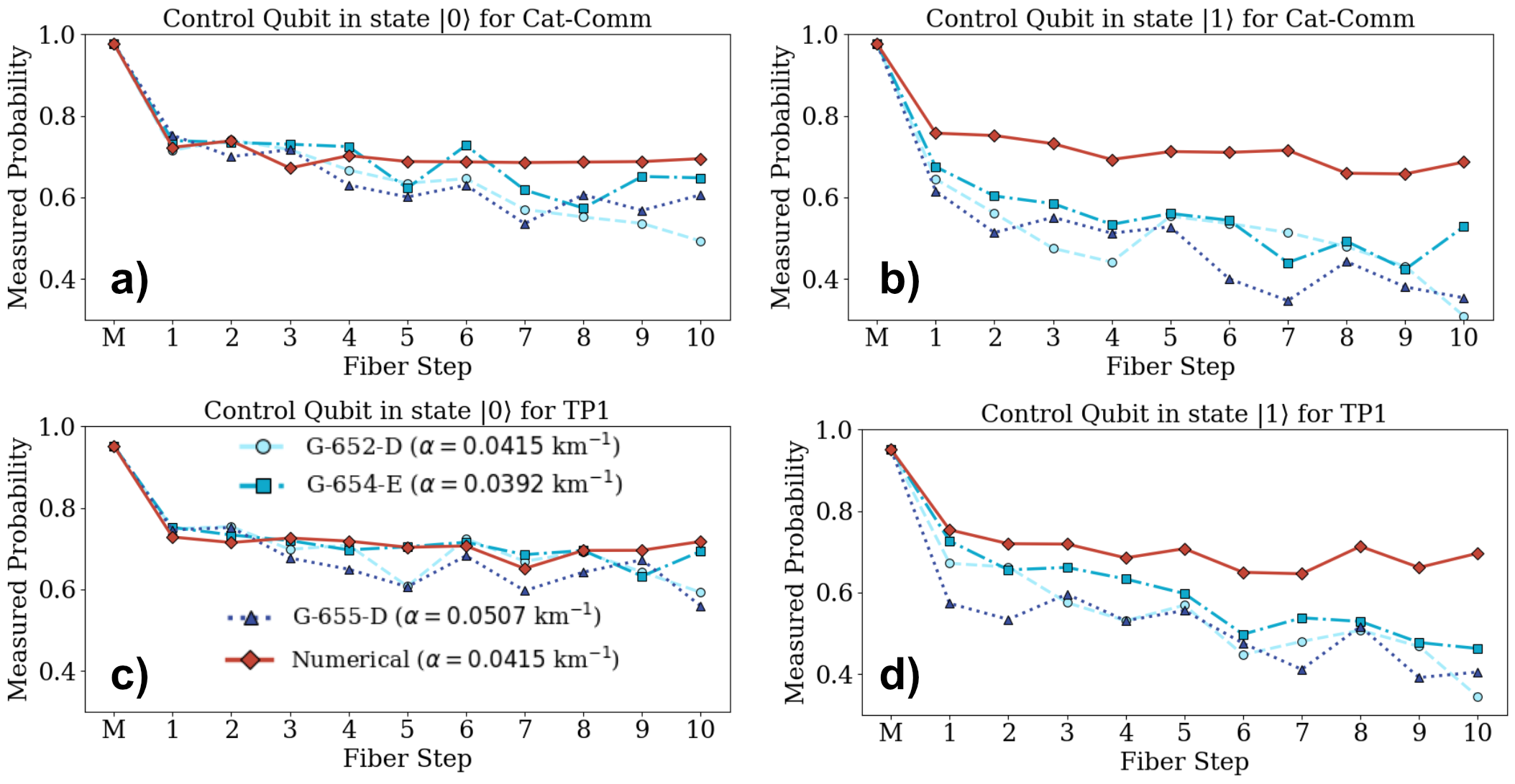}
\caption{The execution results of remote CNOT gates using the \texttt{ibm-torino} backend (with 10,000 shots) and AerSimulator in both monolithic (M) and distributed configurations (1 to 10) (x-axis). The measured probabilities are shown on the y-axis. Each step corresponds to an additional 10\,m fiber segment. Distributed results are shown for G-652-D, G-654-E, and G-655-D fibers and $\kappa_{\mathrm{T}} = 0.5$, together with numerical reference results obtained using Qiskit's \texttt{AerSimulator} with a backend-derived noise model. Panels (a) and (b) correspond to the Cat-Comm protocol with the control qubit initialized in $\ket{0}$ and $\ket{1}$, respectively, while panels (c) and (d) correspond to the TP1 protocol with the control qubit of QPU A initialized in $\ket{0}$ and $\ket{1}$, respectively.}
\label{f:ResultsGates}
\end{figure*}

As shown in Fig.~\ref{f:SimulateGates}(a) and (b), we emulate the noisy RG protocols on a single quantum chip by partitioning two sets of qubits on a single chip as QPUs. In this example, QPU A comprises $q_{1}^{A}$ to $q_{3}^{A}$, and QPU B includes $q_{1}^{B}$ to $q_{4}^{B}$. Our goal is to apply a remote CNOT gate with $q_{1}^{A}$ as the control and $q_{4}^{B}$ as the target qubit. To emulate the noisy communication channel, composed of optical fiber and transducers, we adopt the CM by discretizing the environment into segments, each represented as a qubit.

We use an additional environment qubit in each emulated QPU, labeled $q_{2}^{A}$ and $q_{2}^{B}$, shown as yellow circles in Fig.~\ref{f:SimulateGates}, to represent environmental interactions. These qubits do not correspond to physical qubits in the actual device; instead, they are introduced to emulate the effects of the environment. To visually distinguish them from the real qubits in our diagrams, we depict their circuit lines using dashed lines. We use only one environment qubit per QPU and exploit IBM's \texttt{Reset} operation to recycle it across multiple collision steps. The \texttt{Reset} operation reinitializes the environment qubit to the $\ket{0}$ state before each new interaction. Although its hardware implementation internally involves measurement and conditional operations, the resulting measurement outcomes are neither recorded nor used by the emulation protocol. This reinitialization is also essential for maintaining the Markovian assumption underlying the CM, as it effectively implements the environment refresh implied by the partial trace in Eq.~\ref{eq:dynamics}. By removing any information retained from previous interactions, it ensures that each collision occurs with an independent environment state and prevents memory effects between successive channel segments. Consequently, the protocol does not rely on heralding, post-selection, or protocol-level feed-forward control.

We begin the process by generating entanglement between communication qubits $q_{3}^{A}$ and $q_{1}^{B}$ via a Hadamard gate on $q_{3}^{A}$ followed by a CNOT gate targeting $q_{1}^{B}$. While this would ideally produce high-fidelity entanglement, we deliberately introduce tunable noise (to emulate the errors arising from communication noise) using time-evolution unitary operators $\hat{U}_T$ that simulate interactions with the environment, as defined by the Hamiltonian in Eq.~\ref{Eq:Hamiltonian}. These operators are applied between each communication qubit and its corresponding environment qubit to emulate the effects of transduction noise (with strength $\kappa_{\mathrm{T}} = 0.5$).

After resetting the environment qubit, which only affects the targeted environment qubit and does not have collective effects on other processing qubits, we apply further unitary operators to model the optical fiber segments, each governed by the same Hamiltonian but with coupling $\kappa_{\text{F}}$. The number of fiber steps is variable, providing fine-grained control over the total noise experienced during entanglement distribution.

For comprehensive testing, we emulate the CNOT gate with the control qubit initialized in both $\ket{0}$ and $\ket{1}$. To highlight the gap between theoretical and experimental outcomes, we compare the emulation results from real hardware runs with simulations performed using Qiskit's \texttt{AerSimulator}. 

Fig.~\ref{f:ResultsGates} displays the simulation results. The y-axis indicates the success probability of the CNOT operation, while the x-axis represents the number of fiber steps (i.e., noise collisions). Step 1 corresponds to a single transducer and optical fiber interaction. Although transduction can be further divided into sub-steps with smaller $\kappa_{\mathrm{T}}$, we focus here on the optical channel.
\begin{figure*}[t!]
\centering
\includegraphics[width=1\textwidth]{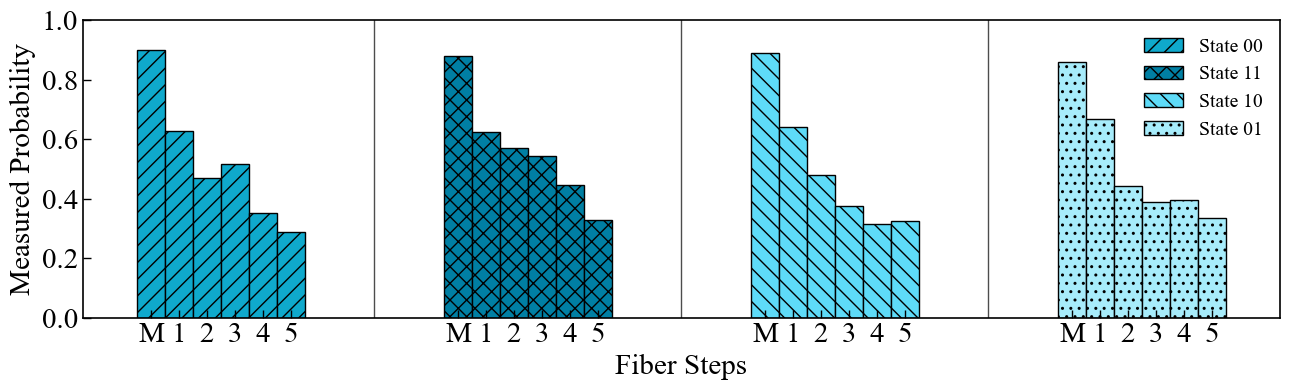}
\captionsetup{width=1\textwidth}
\caption{Comparison of a two-qubit Grover's search algorithm using the \texttt{ibm-torino} backend (with 10,000 shots) in both monolithic (M) and distributed configurations (1 to 5) (x-axis) for the set of marked states $\{00,11,01,10\}$. The measured probabilities of identifying the correct marked state are shown on the y-axis. Each step corresponds to an additional 10\,m fiber segment. Communication is modeled using G-654-E fiber ($\alpha=0.0392$\,km$^{-1}$) and a transduction stage with $\kappa_T=0.5$. The initial step ("1") of the distributed implementation is aligned with the experimental ion-trap results reported in~\cite{main2025distributed}.}
\label{f:GroverResults}
\end{figure*}
In Fig.~\ref{f:ResultsGates}(a) and (b), which correspond to the Cat-Comm CNOT gate with the control initialized in $\ket{0}$ and $\ket{1}$, the red line represents idealized simulation via \texttt{AerSimulator} ($\alpha = 0.0415$ km$^{-1}$), while the light blue, dark blue, and navy blue curves correspond to telecom fiber types G-652-D, G-654-E, and G-655-D, respectively. As expected, the gate success probability decays drastically with increasing steps. Notably, a 30\% initial drop highlights the impact of transduction noise, and a clear performance gap emerges between theoretical simulations and real-hardware results. A similar trend is observed for the TP1 protocol in Fig.~\ref{f:ResultsGates}(c) and (d), corresponding to control states $\ket{0}$ and $\ket{1}$, respectively.

The AerSimulator results in Fig.~\ref{f:ResultsGates} were obtained using a backend-derived noise model, which incorporates calibration data from the target IBM Quantum backend. While this enables the simulator to account for backend-specific estimates of gate and readout errors, agreement with physical hardware is not expected to be exact. The simulator reflects the calibration parameters available when the noise model was generated, whereas hardware execution is subject to the actual device conditions at runtime. Consequently, variations in calibration data, qubit coherence properties, gate fidelities, readout performance, and other hardware effects may result in differences between simulated and experimental outcomes.

Therefore, the discrepancy between the AerSimulator and hardware results presented in Fig.~\ref{f:ResultsGates} should be interpreted as the difference between a calibration-informed simulation and execution on the physical quantum processor. While backend-derived simulations provide a valuable approximation of device behavior, experiments on real quantum hardware remain essential for validating protocol performance under practical operating conditions and for capturing hardware effects that may not be fully represented in calibration-based noise models.
\begin{figure}[t!]
\centering
\includegraphics[width=\columnwidth]{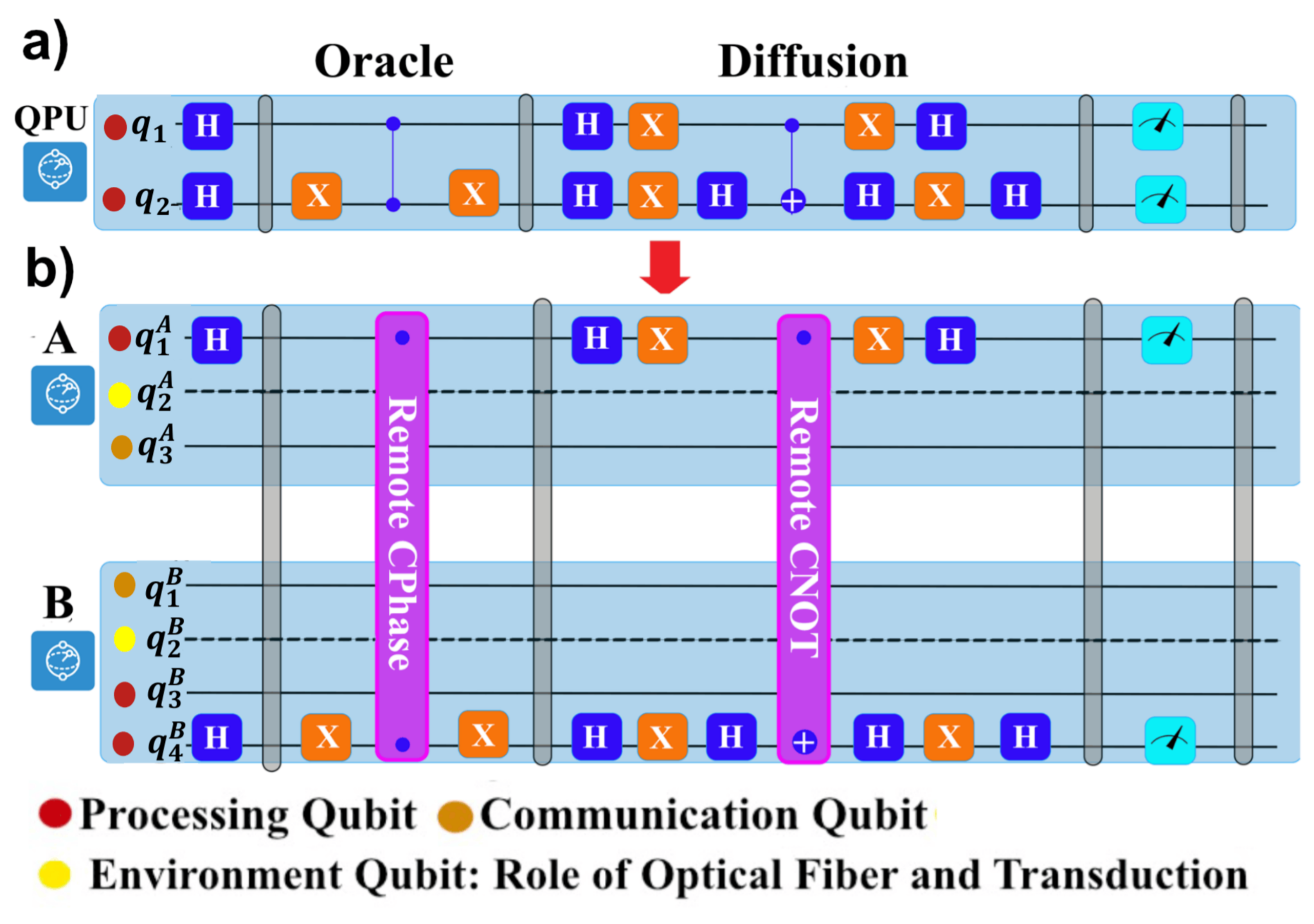}
\caption{The circuits illustrate the execution of Grover's search algorithm for the marked state \texttt{00}. (a) Monolithic implementation of the two-qubit Grover's search algorithm; (b) Distributed implementation suitable for execution across two interconnected QPUs. The qubits $q_{2}^{A}$ and $q_{2}^{B}$, indicated by yellow circles and dashed lines in the circuit, are not native processing or communication qubits of QPU A and QPU B. They are auxiliary environment qubits introduced solely to implement the collision model and emulate the effect of environmental noise on the communication channel.}
\label{f:GroverCircuit}
\end{figure}
As observed from the plots in Fig.~\ref{f:ResultsGates}(b) and (d), the errors of the RGs when the control qubits are in the state $\ket{1}$ appear to be more susceptible to noise. This can be explained by two factors. First, an additional $X$ gate is required to prepare the qubit in the $\ket{1}$ state from the ground state, which introduces extra gate error. Second, decoherence occurs through spontaneous emission of the qubit, corresponding to its $T_1$ relaxation time.

Moreover, the results indicate that TP1 shows slightly better performance than the Cat-Comm protocol. This is because, in TP1, the entire state of the control qubit is teleported, unlike in the cat-based communication scheme. However, it is important to note that teleportation in TP1 collapses the state of the control qubit at its original location. Consequently, if the control qubit is required later in the circuit, its state must be teleported back. Therefore, protocols such as TP1, TP2, and TP-Safe require an additional teleportation step, making them significantly more resource-intensive.

\begin{figure*}[t!]
\centering
\includegraphics[width=1\textwidth]{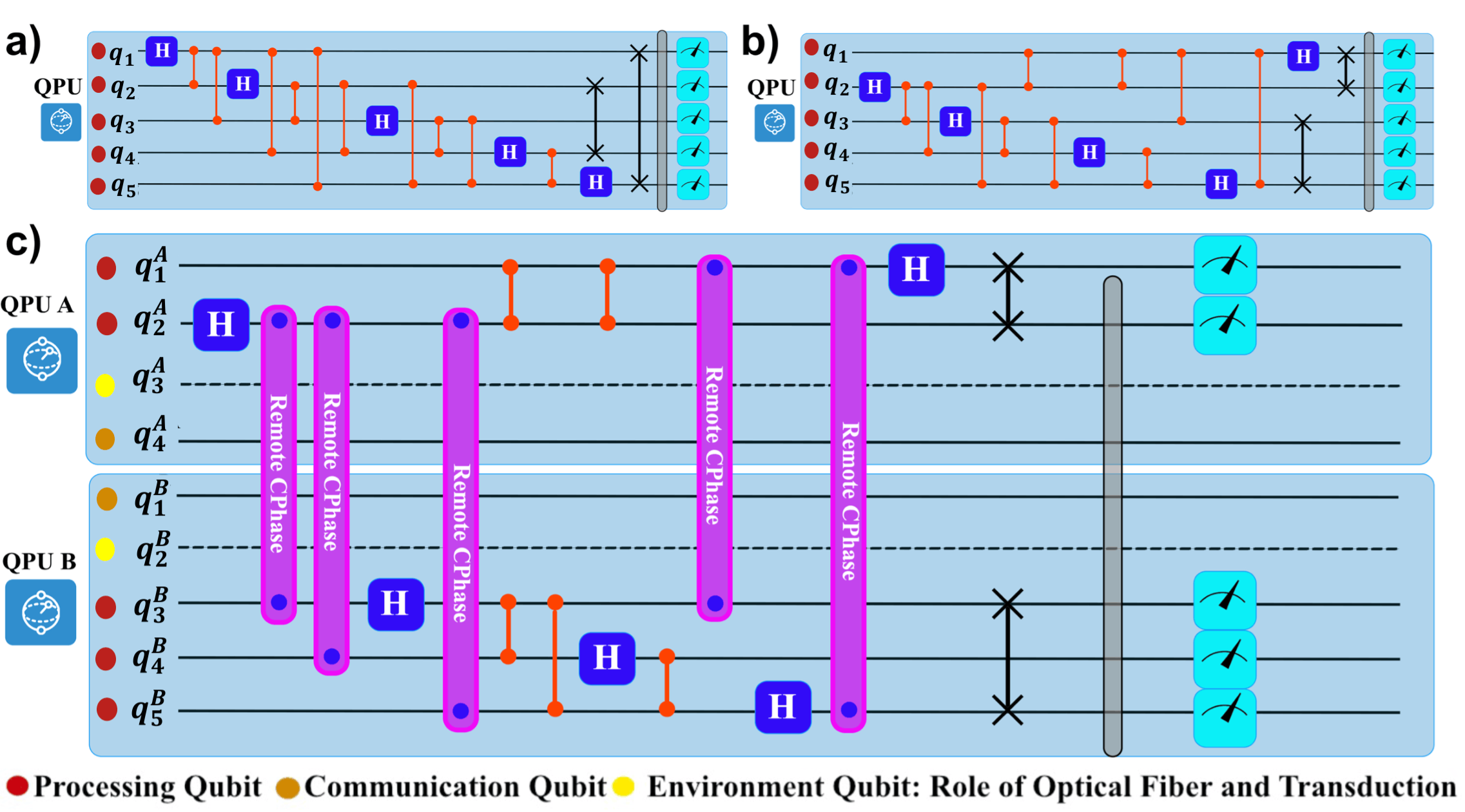}
\captionsetup{width=1\textwidth}
\caption{Execution of the QFT algorithm: (a) Monolithic implementation of the five-qubit QFT; (b) Reconfigured monolithic version optimized to reduce SWAP gates (it must be decomposed into three remote CNOT gates and multiple local gates, which significantly increases the overall error rate); (c) Distributed implementation of the five-qubit QFT across two interconnected QPUs. In this circuit, the qubits $q_{2}^{A}$ and $q_{2}^{B}$, highlighted by yellow circles and dashed lines, are auxiliary environment qubits rather than native qubits of QPU A and QPU B. They are included only to implement the collision model and emulate environmental interactions affecting the communication channel.}
\label{f:QFTCircuit}
\end{figure*}

% ── 3.3 Grover's Search Algorithm ────────────────────────────
\subsection{Grover's Search Algorithm}
\label{sec:Grover}

To further demonstrate the versatility of our framework, we use our emulation framework to implement Grover's search algorithm. Within our QDC emulation framework, we implement Grover's algorithm across two distributed QPUs to locate a marked state. The algorithm consists of repeated applications of the Grover's operator, which includes an oracle that inverts the phase of the marked state and a diffusion operator that amplifies its amplitude~\cite{nielsen2010quantum}.

We focus on the two-qubit version of the algorithm. In the two-qubit case, the algorithm searches over the four computational basis states $\{00, 01, 10, 11\}$. When the target state is $\ket{00}$, the qubits are initialized in $\ket{0}$, and Hadamard gates are applied to create an equal superposition of all states. The oracle then inverts the amplitude of $\ket{00}$, and the diffusion operator reflects all state amplitudes about their mean, boosting the amplitude of the marked state. For two qubits and one marked state, a single iteration of the Grover's operator is sufficient to maximize the probability of measuring the correct result. This setup is illustrated in Fig.~\ref{f:GroverCircuit}(a).

To emulate this on a distributed architecture, we allocate one qubit to each QPU (QPU A and QPU B) and employ CM to introduce communication noise, as described in Section~\ref{sec:SystemModel}. Remote-controlled unitary gates, realized via noisy entanglement between QPUs, are used for inter-qubit operations, and environmental qubits are included in the circuit to enable fine-tuned noise modeling, as shown in Fig.~\ref{f:GroverCircuit}(b). This noise, resulting from transducers and optical fibers, degrades the fidelity of the entangled states and thus reduces the overall success probability of Grover's search algorithm.

We evaluate the algorithm's performance across five optical fiber segments, using a transducer coupling constant of $\kappa_{\mathrm{T}} = 0.5$ and the low-attenuation G-654-E telecom fiber. We test all possible marked states ($00, 01, 10, 11$) and compare the results with both monolithic executions and the experimental data reported for ion-trap-based QPUs~\cite{main2025distributed}. Emulations use the same IBM backend (\texttt{ibm-torino}) and the same shot count as in previous experiments. As expected, the success probability of correctly identifying the marked state decreases as the number of fiber segments increases due to communication noise. As shown in Fig.~\ref{f:GroverResults}, the first-step (indexed as "1") emulation results closely match the experimental findings, where trapped-ion QPUs interconnected via a 2~m single-mode fiber (SMF) achieved a marked-state probability of approximately 71\%~\cite{main2025distributed}. These findings validate our framework's ability to replicate real-world experimental behavior.

% ── 3.4 Quantum Fourier Transform ────────────────────────────
\subsection{Quantum Fourier Transform (QFT)}
\label{sec:QFT}

The QFT is a foundational quantum algorithm used in various applications, including Shor's algorithm for factoring large numbers~\cite{shor1997polynomial}. In our QDC emulation, we implement a five-qubit QFT across two interconnected QPUs, using an approach similar to the one used for the distributed two-qubit Grover's search algorithm. The QFT transforms a quantum state into its Fourier basis, defined for an $N$-dimensional system as $
|j\rangle \mapsto \frac{1}{\sqrt{N}} \sum_{k=0}^{N-1} e^{2\pi i j k / N} |k\rangle,$
where $N=2^n$ for $n$ qubits~\cite{nielsen2010quantum}. As illustrated in Fig.~\ref{f:QFTCircuit}(a), the five-qubit QFT begins with a Hadamard gate applied to the most significant qubit ($q_1$), placing it into superposition. This is followed by a series of controlled phase rotation gates, $R_k$, where the phase angle is $\theta = \pi / 2^{k-1}$, applied between the current qubit and all less significant qubits. This pattern repeats for each subsequent qubit, cascading from the most to the least significant qubit, forming layers of Hadamard and controlled rotations. Finally, a swap layer reverses the qubit order to match the correct output format. In total, the five-qubit QFT circuit requires 5 Hadamard gates, 10 controlled rotation gates, and 2 SWAP gates. Accurate execution depends on precise implementation of small-angle rotations and reliable inter-qubit control, both of which are particularly challenging in noisy environments.

\begin{figure}[t!]
\centering
\includegraphics[width=0.9\columnwidth]{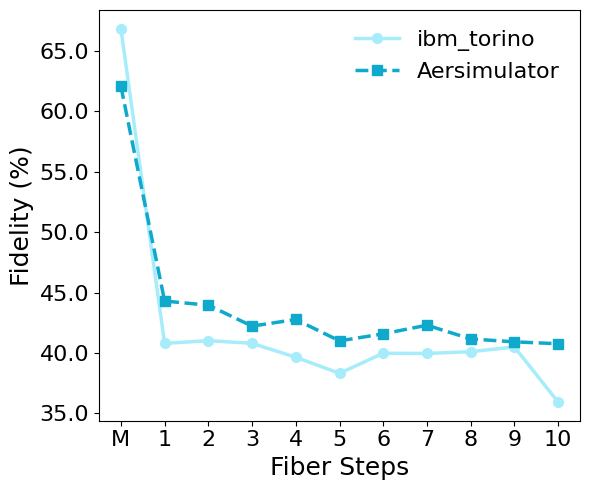}
\caption{The tomography results of a five-qubit QFT algorithm using the \texttt{ibm-torino} backend and AerSimulator in both monolithic (M) and distributed configurations (1 to 10) (x-axis). The y-axis shows the reconstructed-state fidelity, reported as a percentage. Each step corresponds to an additional 10\,m fiber segment. Communication is modeled using G-654-E fiber ($\alpha=0.0392$\,km$^{-1}$) and a transduction stage with $\kappa_T=0.5$. Fidelity is computed using quantum state tomography and evaluated with respect to the ideal QFT output state.}
\label{f:QFTResults}
\end{figure}

To implement this QFT circuit across two QPUs, we distribute qubits across QPU A and QPU B as in the previous section, adding communication qubits and applying RGs. To avoid the SWAP operations in the distributed setting, we instead reorder the qubits in the circuit but retain the original logical ordering for measurement. Specifically, as shown in Fig.~\ref{f:QFTCircuit}(b), we reorder the circuit so that $q_5$ appears first, resulting in the new qubit order: $q_5, q_1, q_2, q_3, q_4$. Nevertheless, we measure the qubits in their original logical order: $q_1, q_2, q_3, q_4, q_5$. In the distributed setting, $q_5$ and $q_1$ are assigned to QPU A, while $q_2, q_3, q_4$ are assigned to QPU B.

Since the QFT outputs a quantum superposition, direct measurement in the computational ($\sigma_z$) basis does not yield meaningful metrics. To address this, we perform quantum state tomography on the processing qubits to reconstruct the output density matrix and compute the fidelity between the ideal and noisy states. Fidelity is a key metric for evaluating quantum operations under noise, which is defined for density matrices $\rho$ and $\sigma$ by the Uhlmann formula~\cite{nielsen2010quantum} $F(\rho, \sigma) = \left( \mathrm{Tr} \left[ \sqrt{ \sqrt{\rho}  \sigma  \sqrt{\rho} } \right] \right)^2,$ in which $\sigma$ is the density matrix of emulation and $\rho$ is the ideally calculated density matrix. This expression generalizes the notion of fidelity to mixed states and reduces to $|\langle \psi | \phi \rangle|^2$ when both states are pure. Fidelity ranges from 0 (completely distinguishable) to 1 (identical), which can also be converted to a percentage, making it an ideal measure for assessing circuit performance in noisy, distributed environments.

For executing the distributed QFT circuit shown in Fig.~\ref{f:QFTCircuit}(c), we use the same simulation settings as in earlier sections, including the transduction parameter $\kappa_{\mathrm{T}}$ and the value of $\kappa_F$ derived from the attenuation coefficient of a G-654-E optical fiber ($\alpha = 0.0392\,\mathrm{km}^{-1}$). For comparison, we also run the monolithic version of the circuit, shown in Fig.~\ref{f:QFTCircuit}(a).

The results presented in Fig.~\ref{f:QFTResults} reveal several important insights into the performance of DQC under realistic communication noise. First, even the monolithic implementation executed on current IBM hardware achieves a fidelity of only approximately $63\%$, indicating that hardware noise alone already contributes significantly to the overall error budget. By comparison, the \texttt{AerSimulator}, configured using a backend-derived noise model, achieves fidelities above $66\%$. Second, the largest reduction in fidelity occurs during the transduction stage, as evidenced by the pronounced initial drop in the figure. In contrast, the additional degradation associated with increasing fiber length is comparatively modest, resulting in a total fidelity reduction to approximately $40\%$. These observations indicate that transduction noise constitutes the dominant communication-related impairment in the considered setting.

The fidelity reduction observed in the distributed implementation should therefore be interpreted in the context of the additional operational complexity required to realize non-local interactions across QPUs. Each non-local controlled-phase operation must be decomposed into a sequence of RG primitives that require the generation and consumption of distributed entanglement, additional gate operations, measurements, and correction steps before the logical operation can be completed. Consequently, the observed degradation reflects the combined effects of transduction-induced noise, communication-induced entanglement degradation, and the additional circuit overhead associated with remote-gate execution. The QFT experiment thus highlights a fundamental challenge of DQC: once non-local operations are introduced, communication becomes a significant contributor to the overall error budget, emphasizing the importance of efficient RG protocols and high-fidelity quantum interconnects.

% ------------------------------------------------------------
%  Section 4 – Conclusion
% ------------------------------------------------------------
\section{Conclusion}
\label{sec:Conclusion}

In this work, we presented a hardware-based framework for emulating DQC architectures using a single quantum processor. By partitioning the coupling map of a quantum device into multiple virtual QPUs, the proposed approach enables the experimental study of distributed quantum protocols without requiring physically interconnected quantum processors.

To capture communication-induced noise between virtual QPUs, we introduced a CM-based framework that emulates the interaction between FQs and their surrounding environment. Because the framework operates directly on physical qubits, it naturally incorporates device-level noise sources and hardware constraints that are difficult to capture in purely classical simulations.

Using IBM superconducting quantum processors as our experimental testbed, we demonstrated remote quantum gate execution under controllable communication noise and implemented distributed versions of Grover's search algorithm and QFT. The Grover's search algorithm experiment reproduces trends reported in recent distributed ion-trap quantum computing experiments, providing additional evidence that the proposed emulator can reproduce experimentally observed behavior in distributed quantum systems. Beyond demonstrating distributed algorithms, the framework enables hardware-level emulation of distributed architectures using currently available quantum devices. Since the implementation relies on circuit-level operations compatible with the Qiskit SDK, the approach can be deployed across different quantum hardware platforms supported by Qiskit, such as IonQ systems.

Overall, the proposed framework provides a practical experimental platform for investigating DQC architectures and communication-induced impairments using present-day quantum hardware. The framework is validated on small, tractable instances; as quantum hardware scales to larger qubit counts, the same approach can be extended to distributed circuits where classical state-vector simulation becomes costly, making hardware-level emulation a feasible experimental testbed.

% ------------------------------------------------------------
%  Data Availability
% ------------------------------------------------------------
\section{Data Availability}

All source code, datasets, and detailed implementation notes for the various components (including a tutorial) of this work are available in the public GitHub repository~\cite{nelyasi_QDC}  and the documentation page~\cite{nelyasi_QDC_Doc}.

% ------------------------------------------------------------
%  Acknowledgments
% ------------------------------------------------------------
\section{Acknowledgments}

We thank Dr. Sima Bahrani for useful and insightful discussions and recommendations on revising the text. This work is supported by the project ``DIGITAL-2022-QCI-02-DEPLOY-NATIONAL'' (Project number: 101113375 --- NQCIS) funded by the EU together with VINNOVA and WACQT. Additional support is provided by VR and GENIE. We thank Chalmers Next Labs for providing a premium account and support for the IBM Quantum Platforms.

% ------------------------------------------------------------
%  Bibliography
% ------------------------------------------------------------
\bibliographystyle{IEEEtran}
\bibliography{sample}

\end{document}